%% file: GRB100621A.tex
\documentclass[traditabstract,longauth]{aa}

\usepackage{amsmath}
\usepackage{amsfonts}
\usepackage{amssymb}
\usepackage{txfonts}
\usepackage{natbib}
\bibpunct{(}{)}{;}{a}{}{,} % to follow the A&A style

\begin{document}
%\linenumbers
\title{Search for TeV Gamma-ray Emission from GRB~100621A, an extremely bright GRB in X-rays, with H.E.S.S.}

\include{paper_authors}

\offprints{D.~Lennarz, \email{dirk.lennarz@gatech.edu}}

\date{Received 05 November 2013 / Accepted 18 March 2014}

\abstract{The long gamma-ray burst (GRB) 100621A, at the time the brightest X-ray transient ever detected by \emph{Swift}-XRT in the $0.3\textrm{--}10$~keV range, has been observed with the H.E.S.S. imaging air Cherenkov telescope array, sensitive to gamma radiation in the very-high-energy (VHE, $>100$~GeV) regime. Due to its relatively small redshift of $z\sim0.5$, the favourable position in the southern sky and the relatively short follow-up time ($<700~\rm{s}$ after the satellite trigger) of the H.E.S.S. observations, this GRB could be within the sensitivity reach of the H.E.S.S. instrument. The analysis of the H.E.S.S. data shows no indication of emission and yields an integral flux upper limit above $\sim$380~GeV of $4.2\times10^{-12}~\rm cm^{-2}s^{-1}$ (95\% confidence level), assuming a simple Band function extension model. A comparison to a spectral-temporal model, normalised to the prompt flux at sub-MeV energies, constraints the existence of a temporally extended and strong additional hard power law, as has been observed in the other bright X-ray GRB~130427A. A comparison between the H.E.S.S. upper limit and the contemporaneous energy output in X-rays constrains the ratio between the X-ray and VHE gamma-ray fluxes to be greater than 0.4. This value is an important quantity for modelling the afterglow and can constrain leptonic emission scenarios, where leptons are responsible for the X-ray emission and might produce VHE gamma rays.}

\authorrunning{H.E.S.S. Collaboration et al.}
\titlerunning{Search for VHE Emission from GRB~100621A with H.E.S.S.}
\maketitle

\section{Introduction}
Gamma-ray bursts (GRBs) are brief flashes of X-ray and soft gamma-ray emission traditionally detected in the keV to MeV energy range \citep[for a review see e.g.][]{GRB_review}. Their origin and internal acceleration mechanisms are among the most enigmatic questions in contemporary astrophysics. Depending on the length of the prompt gamma-ray emission they are divided into two clases, long and short, if their light curves are longer or shorter than 2~s respectively. The prompt emission is in general well described by a Band function \citep{band_function}. GRBs exhibit longer-wavelength afterglows that are important for our understanding of the acceleration mechanisms. The emissions are in general consistent with the fireball model \citep[e.g.][]{Piran_fireball_review}, where the prompt emission is produced by internal shocks originating in the collision of relativistic jets and the afterglow originates from external shocks, when the fireball ejecta run into the surrounding environment.

Observations at higher energies (HE, above $\sim$ 20~MeV) were first made with the Energetic Gamma Ray Experiment Telescope (EGRET). For most detected GRBs the MeV emission is consistent with being a continuation of the GRB spectra at lower energies, without the indication of a cut-off \citep{EGRET_bursts}. However, one GRB exhibited an additional hard power-law component \citep{EGRET_power_law}, which challenges the interpretation in which the HE emission arises from charged particles through synchrotron radiation. There was an indication of temporally-extended emission, most prominently from GRB~940217, where the emission might have lasted more than 5000~s \citep{EGRET_long_burst}.

Today, the Large Area Telescope (LAT) on board the \emph{Fermi Gamma-Ray Space Telescope} (\emph{Fermi}-LAT) observers a subset of very energetic bursts at HE, allowing more detailed studies. For some GRBs \citep[e.g. GRB~080916C,][]{Fermi_LAT_GRB080916C} the emission is consistent with a Band function from keV to GeV energies, whereas other bursts show an additional hard power-law component at HE \citep[e.g. GRB~090902B \& 090510,][]{Fermi_LAT_GRB090902B,Fermi_LAT_GRB090510}, which in some cases exhibits a spectral break \citep[e.g. GRB~090926A,][]{Fermi_LAT_GRB090926A}. These additional spectral components are required for the brightest bursts inside the LAT field of view \citep{Fermi_LAT_GRB_catalogue}. Additionally, \emph{Fermi}-LAT finds that the $>100$ MeV emission of GRBs starts systematically later than the emission at lower energies, reaching delays up to 40~s for GRB~090626 and that the duration is also longer than the low-energy equivalent \citep{Fermi_LAT_GRB_catalogue}. For GRB~130427A, the HE emission lasted 20~h and the observations are not in agreement with being synchrotron radiation in the standard afterglow shock model \citep{Fermi_LAT_GRB130427A}.

GRBs are predicted to emit very-high-energy (VHE, $>100$~GeV) gamma rays in the framework of the fireball model and extending observations of GRBs to the VHE regime is important to further characterise the acceleration and radiation processes at work \citep[e.g. for GRB~130427A, where an inverse Compton scenario has been proposed, see e.g.][]{GRB130427A_Inverse_compton}. Imaging Atmospheric Cherenkov Telescopes (IACTs) such as the High Energy Stereoscopic System (H.E.S.S., see also below), the Major Atmospheric Gamma-Ray Imaging Cherenkov (MAGIC) telescopes and the Very Energetic Radiation Imaging Telescope Array System (VERITAS) are instruments sensitive in this energy range. However, only upper limits on the VHE emission have been reported so far \citep{HESS_GRB,HESS_GRB_PROMPT,MAGIC_GRB,MAGIC_GRB080430,MAGIC_GRB090102,VERITAS_GRB}.

VHE gamma rays are absorbed by interactions with the extragalactic background light (EBL) and can thus only travel limited distances in the Universe \citep[e.g.][]{EBL}. This poses a severe limitation for GRB observations in this energy range since they typically originate from cosmological distances. However, blazar observations show that the level of EBL extinction is lower than previously thought \citep[e.g.][]{HESS_EBL} and it is possible to detect VHE gamma-ray sources even at redshifts above 0.6 with the current generation of telescopes \citep{distance_PKS_1424}.

Its high fluence and very bright afterglow at lower energies identify GRB~100621A as one of the rare and powerful nearby GRBs. Its location within the VHE gamma-ray horizon made this burst a promising target for VHE observations. In this paper, the results of the VHE observations obtained with H.E.S.S. are reported.

\section{GRB~100621A}
GRB~100621A was detected with the Burst Alert Telescope (BAT) on board of the \emph{Swift} satellite \citep{BAT} on June 21, 2010 at 03:03:32 UT \citep{GRB100621A_Swift_BAT_1}, hereafter denoted $t_0$. The duration $T_{90}$, the central time interval of 90\% of the prompt flux detected with BAT between $15\textrm{--}350$~keV, was $(63.6 \pm 1.7)$~s \citep{GRB100621A_Swift_BAT_2} and the burst was located by \emph{Swift's} X-ray Telescope \citep[XRT,][]{XRT} at \mbox{RA(J2000)~=~$\rm 21^h~01^m~13\fs12$} and \mbox{Dec(J2000)~=~$-51\degr$~$06\arcmin$~$22\farcs5$} with an uncertainty of 1.7 arcsec \citep[radius, 90\% confidence level,][]{GRB100621A_Swift_XRT_1}. This burst featured an extremely bright X-ray afterglow \citep{GRB100621A_Swift_XRT_2}, making it the brightest X-ray transient ever detected by the XRT at that time. Recently, GRB~100621A has been surpassed by GRB~130427A \citep{Swift_GRB130427A}.

The Konus-W experiment \citep{Konus-Wind} on board the \emph{WIND} spacecraft (Konus-\emph{Wind}) detected a fluence of GRB~100621A in the energy range of 20~keV$\textrm{--}$2~MeV of $(3.6\pm0.4)\times 10^{-5}$~erg/cm$^2$ within 74 seconds after the trigger \citep{GRB100621A_KonusWind}. The time-integrated spectrum of the burst is best fit by a Band function \citep{band_function}, where the low-energy photon index is $-1.69_{-0.07}^{+0.08}$, the high-energy photon index is $-2.46_{-0.45}^{+0.13}$ and the peak energy of the spectral energy distribution is $E_\mathrm{p}=95_{-8}^{+9}$ keV \citep[quoted errors at the 68\% confidence level,][]{GRB100621A_KonusWind_private}. The break energy $E_0$ is directly related to $E_p$ via: $E_0 = E_\mathrm{p}/(\alpha+2)$ and the normalisation constant of the Band function can be calculated in such a way that the fluence corresponds to the one measured by Konus-\emph{Wind}.

%\begin{align}\label{eq:band_function}
%  N(E) = \left\{
%  \begin{array}{l l}
%    N_0 \left(\frac{E}{100~\rm{keV}}\right)^\alpha\exp\left(-\frac{E}{E_0}\right), & \quad \text{for }(\alpha-\beta)E_0\geq E,\\
%    N_1 \left(\frac{E}{100~\rm{keV}}\right)^\beta, & \quad \text{for }(\alpha-\beta)E_0\leq E,\\
%  \end{array} \right.
%  %\left(\frac{(\alpha-\beta)*E_0}{E_{\ref}}\right)^{(\alpha-\beta)}\exp\left(\beta-\alpha\right)
%\end{align}
%with the low-energy photon index $\alpha=-1.69_{-0.07}^{+0.08}$, high-energy photon index $\beta = -2.46_{-0.45}^{+0.13}$ and peak energy of the spectral energy distribution $E_\mathrm{p}=95_{-8}^{+9}$ keV \citep{GRB100621A_KonusWind_private}, where $E_\mathrm{p}$ is directly related to $E_0$ via: $E_0 = E_\mathrm{p}/(\alpha+2)$ and the normalisation constants $N_1=N_0\left[\frac{(\alpha-\beta)E_0}{100~\rm{keV}}\right]\exp\left(\beta-\alpha\right)$ can be calculated from the fluence. The quoted errors are at the 68\% confidence level.

The redshift of GRB~100621A has been measured to be $z=0.542$ with the Very Large Telescope (VLT) and the X-shooter spectrograph \citep{GRB100621A_redshift_1}. This value was derived from bright emission lines of the host galaxy. The GRB afterglow shows extreme reddening, which is in strong contrast to the blue host galaxy. This suggests that the immediate GRB environment is more dusty than the rest of the host galaxy \citep{GRB100621A_redshift_2}. The optical/near-infrared afterglow exhibits a complex temporal evolution with a steep increase in brightness from around 3.5 to 4.5~ks after the trigger \citep{GRB100621A_redshift_2}.

The GRB position was not visible for the \emph{Fermi} spacecraft at the time of the \emph{Swift} trigger due to occultation by the Earth. There is also no LAT coverage of the burst position during the H.E.S.S. observations (see below).

The detection prospects of GRB~100621A in the VHE regime are hard to estimate from the prompt spectrum, because observations carried out by Cherenkov telescopes are typically not contemporaneous with the satellite-based observations, but start on the order of 100~s later. One can, motivated by the unbroken spectra seen by \emph{Fermi}-LAT for some bursts and neglecting a possible spectral cut-off and time delay, extrapolate the prompt, time-integrated spectrum measured by Konus-\emph{Wind} to the VHE regime (Band function extension model). The effect of the absorption on the EBL is estimated using a model by \cite{Franceschini_EBL}, which is interpolated to the GRB redshift. Given the used assumptions this extrapolated flux ($6.2\times10^{-14}~\rm cm^{-2}s^{-1}TeV^{-1}$ at 1~TeV) is in reach of the H.E.S.S. instrument.

The temporal evolution of the Band function extension model flux can be modelled for example as in \cite{GRBs_with_CTA}, assuming that the flux in the VHE regime is constant during $T_{90}$ and then decays as a power law $\left(\frac{t}{T_{90}}\right)^{-\gamma}$ when the delay $t$ to the prompt emission grows. This model, consisting of the Band function extension model, EBL absorption and the temporal decay, constitutes the spectral-temporal model used in the analysis, assuming $\gamma=1.5$.

In the spectral-temporal model the flux estimation will be below the reach of the H.E.S.S. instrument for typical observational delays. However, since other bright GRBs seen by \emph{Fermi}-LAT, like e.g. GRB~130427A, exhibit an additional hard power-law component, one can speculate on temporally-extended and delayed HE emission here. If the component seen in GRB~130427A extended to slightly higher energies than the highest energy photon observed, it would be easily detectable at VHE. It is however unclear, if such a component exists in GRB~100621A, if it extrapolates to the VHE regime and which spectral shape or flux level it should have at the time of observation. Nevertheless, the H.E.S.S. observations provide the ability to detect a possible temporally-extended and strong VHE emission from a hard power-law component.

\section{The High Energy Stereoscopic System}
H.E.S.S. is an array of four IACTs located 1800~m above sea level in the Khomas Highland of Namibia. It is sensitive to VHE gamma rays between hundreds of GeV to tens of TeV by detecting Cherenkov light emitted when the gamma ray is absorbed in the atmosphere in an extensive air shower. Such observations are taken during the parts of the nights without any moon and no clouds in the field of view. Each telescope has a 13~m diameter and $\sim100~\rm{m^2}$ tessellated mirror surface arranged in a Davies-Cotton design with a focal length of 15~m. The telescopes are arranged in a square with 120~m side length with one diagonal oriented north-south. Furthermore, each telescope is equipped with a pixelated camera of 960 photomultiplier tubes (PMTs) with Winston cones in front to improve the light collection efficiency. One pixel subtends approximately $0.16\degr$, resulting in a total field of view of $5\degr$ in diameter. The triggering is done in three different stages: at PMT level, at telescope and at array level \citep{trigger_paper}. Only events recorded by at least two of the four telescopes are used, allowing stereoscopic image analysis. This results in an angular resolution (68\% containment) of typically $0.1\degr$ and an energy resolution of $\sim15\%$. The H.E.S.S. effective area and energy threshold are largely influenced by the zenith angle of the observation, leading to a higher energy threshold the larger the zenith angle of the observation. A more comprehensive summary of H.E.S.S. can be found in \citet{HESS_crab} and the references therein.

\section{Data collection and analysis}
In order to allow rapid follow-up observations, the H.E.S.S. data acquisition system is connected to the GRB Coordinates Network (GCN)\footnote{http://gcn.gscfc.nasa.gov}. Notices of GRBs detected by satellites are received via socket connection and automatically processed on site. Currently, H.E.S.S. accepts notices from \emph{Swift}-BAT and \emph{Fermi}-LAT as triggers if they have a positional uncertainty $<2.5\degr$ and more detailed trigger conditions are met e.g. the significance, a position incompatible with known sources and the quality of the trigger data. Observations should be started immediately by the observers present at the telescopes if the trigger is received during dark time (i.e. night and no moon) with fair weather conditions and if the GRB position can be observed with a zenith angle smaller than $45\degr$ to ensure a reasonably low energy threshold. Recently, this human-in-the-loop process has been replaced by a fully automated repointing procedure, which was however not yet present at the observation of GRB~100621A. Further technical details of the H.E.S.S. GRB programme can be found in \cite{HESS_GRB_technical}.

The trigger for GRB~100621A from \emph{Swift}-BAT was received in Namibia at 03:04:01 UT, which is 29~s after $t_0$. However, due to technical problems, observations were started only at 03:14:55 UT which is 683~s after $t_0$. Due to moonrise only two observations with a nominal duration of 28~min were taken. The burst was observed in ``wobble mode'' in which the observation position is displaced from the centre of the camera \citep{WOBBLE} to allow for observation and background estimation from the same field of view \citep[reflected-region-background model, see][]{BG_paper}. The first observation was displaced by $-0.5\degr$ in declination and started at a zenith angle of $31.7\degr$, reaching a final position of $34.6\degr$ (mean value of $32.7\degr$) with a deadtime-corrected livetime of 1576~s. For the second observation (displaced $0.5\degr$ in declination), the zenith angle range was $34.1\textrm{--}37.3\degr$ (mean value of $36.1\degr$). It started at 03:45:23 UT and had a livetime of 1574~s. All data were taken during good weather conditions with good hardware status of all four telescopes.

The data calibration, image cleaning, Hillas moment calculation \citep{HILLAS} and event reconstruction is done as described in \cite{HESS_crab} with the standard H.E.S.S. analysis software\footnote{version hap-11-07-pl01}. In this reference, three different selection cuts (standard, hard, loose) to reject background caused by cosmic-ray showers are described, suited for different source scenarios. The background rejection can also be done with a multivariate cut using a decision tree obtained from a boosting algorithm \citep[boosted decision tree, for details see][]{TMVA}. Recently, selection cuts corresponding to the loose cuts from \cite{HESS_crab} have been added to the multivariate analysis. A size-cut of 40 photo electrons, a $\theta^2$-cut of 0.02~degrees$^2$ (where $\theta$ is the angular distance between the reconstructed event direction and the assumed source position), and a $\zeta$-cut of 0.85 \citep[where $\zeta$ denotes the classifier of the boosted decision tree, see][]{TMVA} are used. Due to EBL absorption, the spectrum {of GRB~100621A is expected to be very soft, which makes the sensitivity of the analysis highly dependent on the energy threshold. The lower intensity cut of loose cuts reduces the energy threshold compared to standard and hard cuts. Thus, the multivariate loose cuts have the highest sensitivity and are used in this analysis.

After applying the selection cuts, the number of events ($N_{\rm{on}}$) in the signal region (``on-region'') around the GRB position and the number of events ($N_{\rm{off}}$) in the regions used to estimate the background (``off-regions'') can be used to calculate the significance of the gamma-ray excess using Eq. (17) of \citet{LiMa}. A normalisation factor $\alpha$ is applied to correct for the different number of on- and off-regions.

The energy threshold for the spectral analysis, $E_{\rm th}=$ 383~GeV, is defined by the energy below which the energy bias becomes larger than 10\%. This approach is conservative, because it reduces systematic uncertainties in the estimation of the effective area. H.E.S.S. can still detect gamma rays with energies below this value and all events are used when estimating the significance. However, the spectral analysis is restricted to events with reconstructed energies above the energy threshold.

\section{Results}
\begin{table}[t]
\centering
\caption{Results of the search for excess photons.}
\label{tab:results_significance}
\renewcommand{\arraystretch}{1.3}
\begin{tabular}{lllllll}
  \hline
  \hline
  & N$_{\rm{on}}$ & N$_{\rm{off}}$ & $\alpha$ & $N_{\rm excess}$ & Significance\\
  \hline
  Total                    & 46 & 427 & 0.118 & $-4^{+8}_{-7}$ & $-0.6$ \\
  First 300~s              &  8 &  39 & 0.125 & $ 3^{+3}_{-3}$ &  $1.2$ \\
  $1^{\rm st}$ observation & 26 & 197 & 0.125 & $ 1^{+6}_{-5}$ &  $0.3$ \\
  $2^{\rm nd}$ observation & 20 & 230 & 0.111 & $-6^{+5}_{-5}$ & $-1.1$ \\
  \hline
\end{tabular}
\tablefoot{$N_{\rm{on}}$ is the number of gamma-ray candidates in the signal region around the GRB position and $N_{\rm{off}}$ the background estimate. When scaled by the normalisation factor $\alpha$ they yield the number of excess events $N_{\rm excess} = N_{\rm{on}}-\alpha N_{\rm{off}}$.}
\end{table}

The results of the analysis of the H.E.S.S. data taken for GRB~100621A are shown in Table~\ref{tab:results_significance}. No excess is observed using the total data set. In order to search for emission on shorter time scales and closer to $t_0$ a further analysis was done on each observation separately and on the events corresponding to the first 300~s of the first observation. Shorter time scales are not possible because the number of events in the on-region would become too low to estimate the significance. No significant excess is found here either. The result for the total dataset has also been crosschecked with an independent calibration and analysis of the data \citep{Paris_MVA}.

\begin{table}[t]
\centering
\caption{Integral flux upper limits.}
\label{tab:results_flux}
\begin{tabular}{c|c|c|c}
  \hline
  \hline
  & Above $E_{\rm th}$\tablefootmark{a} & \multicolumn{2}{|c}{Differential\tablefootmark{b} at}\\
  \hline
                           &                     & $E_{\rm th}$        & 1~TeV  \\
  \hline
  Total                    & $4.2\times10^{-12}$ & $6.1\times10^{-11}$ & $1.0\times10^{-13}$ \\
  $1^{\rm st}$ observation & $6.4\times10^{-12} $& $9.4\times10^{-11}$ & $1.5\times10^{-13}$ \\
  $2^{\rm nd}$ observation & $3.8\times10^{-12} $& $5.3\times10^{-11}$ & $1.6\times10^{-13}$ \\
  \hline
\end{tabular}

\tablefoot{
Upper limits correspond to a confidence level of 95\% as derived from the H.E.S.S. spectral analysis, assuming the EBL absorbed simple Band function extension model. For the first observation and the total data set the energy threshold is $E_{\rm th}=$ 383~GeV and for the second observation $E_{\rm th}=$ 422~GeV. The integral upper limits are also expressed as a differential flux at certain energies.
\tablefoottext{a}{Units $\rm cm^{-2}s^{-1}$}
\tablefoottext{b}{Units $\rm cm^{-2}s^{-1}TeV^{-1}$}
}
\end{table}

\begin{figure}[t]
\includegraphics*[width=\columnwidth]{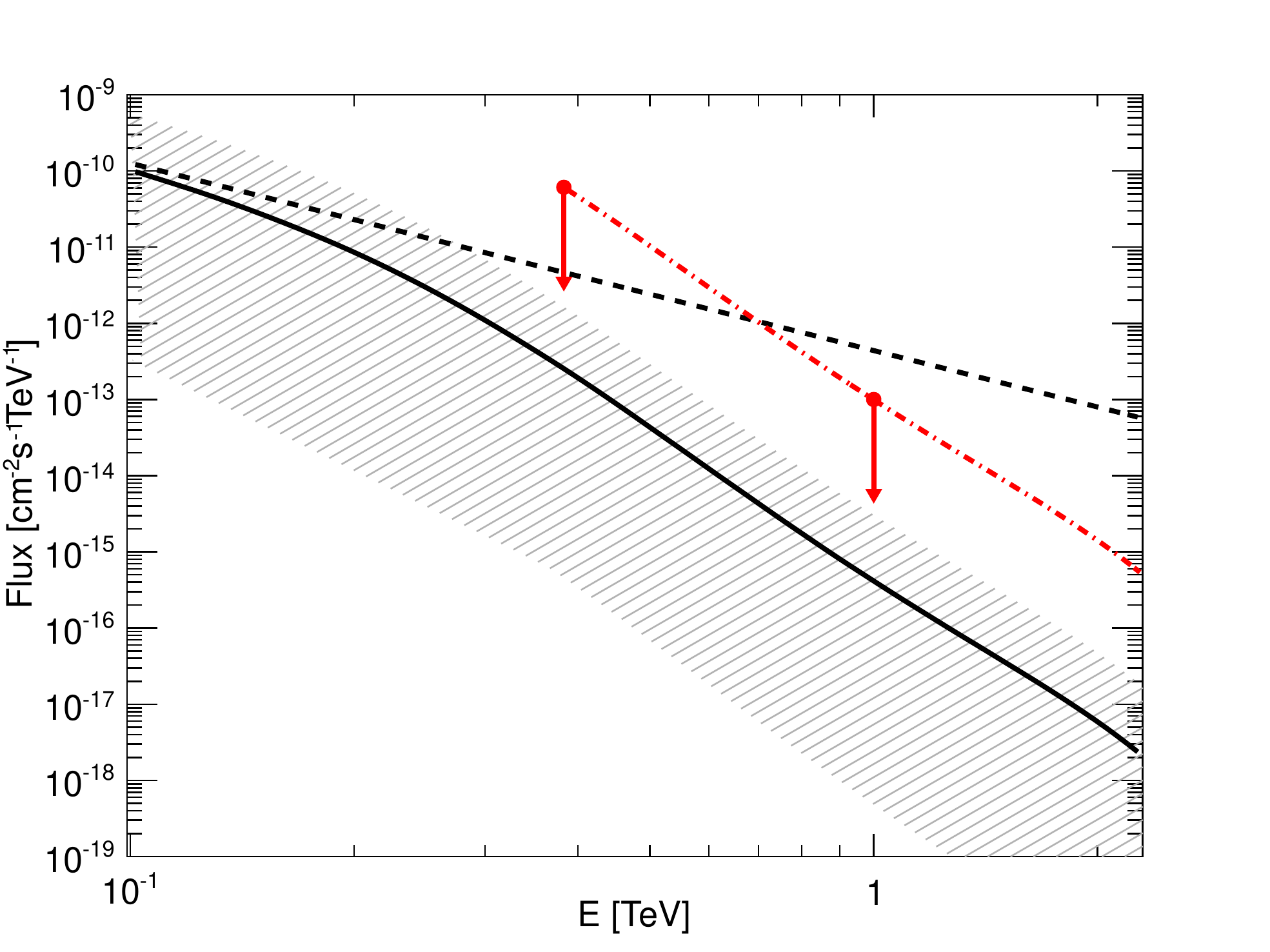}
\caption{The solid line shows the spectral-temporal model matching the H.E.S.S. observation window, while the dashed line shows the same spectrum without applying the EBL model by \cite{Franceschini_EBL}. It can be seen that the spectral shape is dominated by the EBL absorption in the H.E.S.S. energy range. The red dashed-dotted line shows the spectrum that corresponds to the limits given in Table~\ref{tab:results_flux} as obtained by the analysis of the total data set, where the red dots are the two given differential representations. The shaded area shows the effect of varying the Konus-\emph{Wind} high-energy photon index $\beta$ within its one-sigma error.}
\label{fig:bandfunction_EBL_limit}
\end{figure}

Upper limits on the number of excess events are calculated using the method of \cite{Rolke}. These upper limits are converted to integral flux upper limits using the H.E.S.S. effective area. The spectral shape is assumed to follow the Band function extension model plus EBL absorption (a temporal component plays no roll in the calculation). The integral limit can be presented as a differential flux on the assumed spectrum of $1.0\times10^{-13}~\rm cm^{-2}s^{-1}TeV^{-1}$ at 1~TeV at 95\% confidence level (see Table~\ref{tab:results_flux}).

Figure~\ref{fig:bandfunction_EBL_limit} shows a graphical representation of the upper limit and compares it to the spectral-temporal model. It can also be seen that the spectral shape in the H.E.S.S. energy range is mostly dominated by the EBL absorption. Thus, changing the spectral model from the Band function extension model to e.g. an $E^{-2}$ spectrum would change the limits only marginally. Changing the decay factor $\gamma$ in the temporal decay e.g. to 1.0 would move the model up by a factor of $\sim5$, which is small compared to the other uncertainties of the extrapolation. This decay index has been observed by \emph{Fermi}-LAT, however the characteristic time scale is the time of the LAT peak emission \citep{Fermi_LAT_GRB_catalogue} and its relation to the $T_{90}$ at lower energies remains unclear.

\begin{figure}[t]
\includegraphics*[width=\columnwidth]{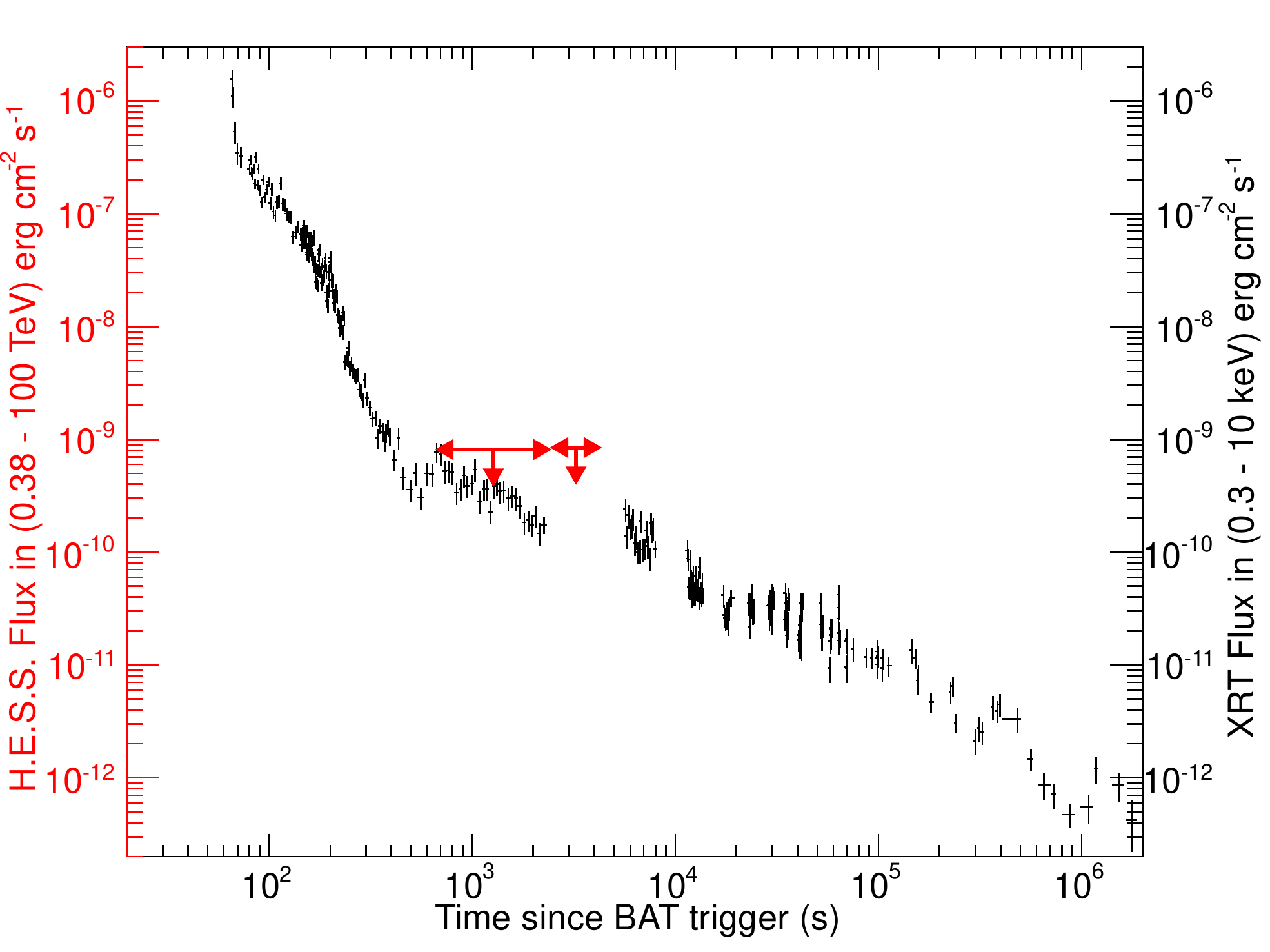}
\caption{Comparison of the VHE upper limits (95\% confidence level) on the energy output above the energy threshold (in lighter colour) using the Band function extension model (no EBL correction applied) with the XRT energy flux \citep[in darker colour, de-absorbed, from the Swift Burst Analyser,][]{Swift_1, Swift_2}. Horizontal arrows indicate the start and end time of the observations from which the corresponding upper limit is derived.}
\label{fig:bandfunction_lightcurve}
\end{figure}

In Fig.~\ref{fig:bandfunction_lightcurve} the energy output after correcting for absorption effects in the H.E.S.S. (0.38 - 100~TeV) and XRT energy range (0.3 - 10~keV) is compared. As can be seen, GRB~100621A exhibited an extremely bright X-ray afterglow at earlier times. The H.E.S.S. observations were obtained during the shallow X-ray phase and do not cover the steep increase in brightness in the optical/near-infrared afterglow. During the first observation the ratio between the energy output (using the energy ranges given in Fig.~\ref{fig:bandfunction_lightcurve}) in X-ray ($F_{\rm X-ray}$) and VHE ($F_{\rm VHE}$) can be constrained to be $\frac{F_{\rm X-ray}}{F_{\rm VHE}} > 0.4$ at 95\% confidence level.
%\begin{align}\label{eq:ratio}
%  \frac{F_{\rm X-ray}}{F_{\rm VHE}} > 0.4
%\end{align}

\section{Interpretation}
Figure~\ref{fig:bandfunction_EBL_limit} shows that the H.E.S.S. upper limit is above the simple temporal Band function extension model. Furthermore, to illustrate the large uncertainty that comes with the Band function extension model the Konus-\emph{Wind} high-energy photon index is varied within its one-sigma error. This gives an uncertainty of several orders of magnitude, without taking into account the errors of the other model parameters and their correlations with $\beta$. Thus, the H.E.S.S. upper limit is not able to exclude the simple temporal Band function extension model.

A temporally-extended and additional hard power law with an $E^{-2}$ spectrum would have been detected by H.E.S.S. if its unabsorbed fluence between 383~GeV and 10~TeV is in excess of $\sim3\times10^{-6}\rm~erg~cm^{-2}$. The contribution from the temporal Band function extension model is small at the time of the H.E.S.S. observations and can be neglected. This fluence limit is within a factor of about $\pm2$ independent of the power law index (between -1.5 and -2.5), because the spectral shape in the H.E.S.S. energy range is dominated by the imprint of the EBL absorption.

Variations of the spectral index have however a strong influence on the fluence at lower energies. The above limit (for $E^{-2}$) corresponds to $1\times10^{-5}\rm~erg~cm^{-2}$ or $2\times10^{-6}\rm~erg~cm^{-2}$ between 10~keV and 10~GeV and $>100~\rm MeV$ respectively, while for $E^{-1.5}$ it changes to $2\times10^{-7}\rm~erg~cm^{-2}$ and $4\times10^{-8}\rm~erg~cm^{-2}$. For GRB~130427A, the LAT measured a $>100~\rm MeV$ fluence in the 100~ks following the trigger of $(7\pm 1)\times10^{-4}\rm~erg~cm^{-2}$ with a typical spectral index of $E^{-2}$\citep{Fermi_LAT_GRB130427A}. The presence of a component that strong during the afterglow phase of GRB~100621A can be excluded, which is remarkable, given that both GRBs were of similar brightness in X-rays.

Motivated by the temporal model discussed earlier, one can assume that the fluence during the H.E.S.S. observations is at the 1\% level compared to the prompt phase. Thus, an additional component as strong as in GRB~090902B \citep[$1.007^{+0.059}_{-0.057}\times10^{-4}\rm~erg~cm^{-2}$ between 10~keV and 10~GeV, obtained during the time of the first LAT photon and the GBM $T_{95}$,][]{Fermi_LAT_GRB_catalogue} is excluded if it had a spectrum following $E^{-1.5}$ at the time of the H.E.S.S. observations. Both interpretations are subjected to the assumption of no spectral break in the extra component.

In a leptonic scenario, the X-ray afterglow is typically modelled as electron synchrotron emission in the external shock. The accelerated electrons could upscatter photons generated e.g. by synchrotron emission from the same population of electrons via the inverse Compton process (synchrotron self-Compton, SSC), which would lead to VHE radiation. In most modeles the energy outputs in X-rays and VHE gamma radiation are proportional. Thus the upper limits on the energy output obtained here can be used to constrain such modelling.

%Hadronic models are mostly applied to the prompt and early afterglow phase and might explain the additional hard power-law component seen by \emph{Fermi}-LAT in some bursts by accelerated protons. It is however unclear what their contribution to a possible VHE signal could be at the time scale of the H.E.S.S. observations. The limits derived here could also be used to constrain hadronic models.

\section{Summary}
In this paper, the analysis of the H.E.S.S. data on GRB~100621A is presented. A significant excess has neither been observed in the total data set, nor on shorter time scales closer to the prompt emission. This constrains the possibility of a temporally-extended emission in the form of an additional hard power law like it has been observed by \emph{Fermi}-LAT in previous bright bursts. A component as strong as in GRB~130427A is not compatible with the H.E.S.S. measurements.

GRB~100621A is one of the brightest X-ray sources detected by \emph{Swift} with a very bright X-ray afterglow. The H.E.S.S. observations started during the shallow decline of the X-ray light curve and the upper limits on the energy output during that time are comparable to the level of the X-ray emission. The ratio between the X-ray and VHE flux is constrained to be greater than 0.4, which can constrain the synchrotron modelling of the afterglow due to the apparent lack of detected inverse Compton emission.

The advent of H.E.S.S. II, which is the world's largest Imaging Atmospheric Cherenkov Telescope, significantly enhances the chances of a VHE GRB detection. The telescope will have a lower energy threshold (tens of GeV) and a higher performance drive system that will reduce the response time to a GRB alert \citep{HESS_II}.

\begin{acknowledgements}
The support of the Namibian authorities and of the University of Namibia in facilitating the construction and operation of H.E.S.S. is gratefully
acknowledged, as is the support by the German Ministry for Education and Research (BMBF), the Max Planck Society, the German Research Foundation (DFG), the French Ministry for Research, the CNRS-IN2P3 and the Astroparticle Interdisciplinary Programme of the CNRS, the U.K. Science and Technology Facilities Council (STFC), the IPNP of the Charles University, the Czech Science Foundation, the Polish Ministry of Science and  Higher Education, the South African Department of Science and Technology and National Research Foundation, and by the University of Namibia. We appreciate the excellent work of the technical support staff in Berlin, Durham, Hamburg, Heidelberg, Palaiseau, Paris, Saclay, and in Namibia in the construction and operation of the equipment.\end{acknowledgements}

\bibliographystyle{aa}
\bibliography{references}

\end{document}

%% file: paper_authors.tex
\author{H.E.S.S. Collaboration
\and A.~Abramowski \inst{1}
\and F.~Aharonian \inst{2,3,4}
\and F.~Ait Benkhali \inst{2}
\and A.G.~Akhperjanian \inst{5,4}
\and E.~Ang\"uner \inst{6}
\and G.~Anton \inst{7}
\and S.~Balenderan \inst{8}
\and A.~Balzer \inst{9,10}
\and A.~Barnacka \inst{11}
\and Y.~Becherini \inst{12}
% ...
}

\institute{
Universit\"at Hamburg, Institut f\"ur Experimentalphysik, Luruper Chaussee 149, D 22761 Hamburg, Germany \and
Max-Planck-Institut f\"ur Kernphysik, P.O. Box 103980, D 69029 Heidelberg, Germany \and
Dublin Institute for Advanced Studies, 31 Fitzwilliam Place, Dublin 2, Ireland \and
National Academy of Sciences of the Republic of Armenia, Yerevan  \and
Yerevan Physics Institute, 2 Alikhanian Brothers St., 375036 Yerevan, Armenia \and
Institut f\"ur Physik, Humboldt-Universit\"at zu Berlin, Newtonstr. 15, D 12489 Berlin, Germany \and
Universit\"at Erlangen-N\"urnberg, Physikalisches Institut, Erwin-Rommel-Str. 1, D 91058 Erlangen, Germany \and
University of Durham, Department of Physics, South Road, Durham DH1 3LE, U.K. \and
DESY, D-15735 Zeuthen, Germany \and
Institut f\"ur Physik und Astronomie, Universit\"at Potsdam,  Karl-Liebknecht-Strasse 24/25, D 14476 Potsdam, Germany \and
Nicolaus Copernicus Astronomical Center, ul. Bartycka 18, 00-716 Warsaw, Poland \and
Landessternwarte, Universit\"at Heidelberg, K\"onigstuhl, D 69117 Heidelberg, Germany
% ...
}